# From Graphene to Bismuth Telluride: Mechanical Exfoliation of Quasi-2D Crystals for Applications in Thermoelectrics and Topological Insulators

Desalegne Teweldebrhan, Vivek Goyal and Alexander A. Balandin*

*Nano-Device Laboratory, Department of Electrical Engineering and Materials Science and Engineering Program, Bourns College of Engineering, University of California – Riverside, Riverside, California 92521 USA*

(Submitted to *Nano Letters* in October 2009, to appear in March 2010)

**Abstract**

Bismuth telluride ($Bi_2Te_3$) and its alloys are the best bulk thermoelectric materials known today. The stacked quasi-two-dimensional (2D) layers of $Bi_2Te_3$ were also identified as topological insulators. In this paper we describe a method for "graphene-inspired" exfoliation of crystalline bismuth telluride films with a thickness of a few atoms. The atomically thin films were suspended across trenches in $Si/SiO_2$ substrates, and subjected to detail characterization. The presence of the *van der Waals gaps* allowed us to disassemble $Bi_2Te_3$ crystal into its *quintuple* building blocks – five mono-atomic sheets – consisting of $Te^{(1)}$-Bi-$Te^{(2)}$-Bi-$Te^{(1)}$. By altering the thickness and sequence of atomic planes we were able to create "designer" non-stoichiometric quasi-2D crystalline films, change their composition and doping, as well as other properties. The exfoliated *quintuples* and ultra-thin films have low thermal conductivity, high electrical conductivity and enhanced thermoelectric properties. The obtained results pave the way for producing stacks of crystalline bismuth telluride quantum wells with the strong spatial confinement of charge carriers and acoustic phonons for thermoelectric devices. The developed technology for producing free-standing quasi-2D layers of $Te^{(1)}$-Bi-$Te^{(2)}$-Bi-$Te^{(1)}$ creates an impetus for investigation of the topological insulators and their possible practical applications.

* Corresponding author; electronic address: balandin@ee.ucr.edu ; group web-site: http://ndl.ee.ucr.edu

Bismuth telluride ($Bi_2Te_3$) is a unique material with a potential for diverse range of applications. Since the discovery of its extraordinary thermoelectric properties, $Bi_2Te_3$ has become a vital component for thermoelectric industry [1-3]. Bulk $Bi_2Te_3$-based materials are known to have the highest thermoelectric figure of merit, $ZT \sim 1.14$ at room temperature (RT). The thermoelectric figure of merit is defined as $ZT=S^2\sigma T/K$, where $S=-\Delta V/\Delta T$ is the Seebeck coefficient ($\Delta V$ is the voltage difference corresponding to a given temperature difference $\Delta T$), $\sigma$ is the electrical conductivity and $K$ is the thermal conductivity, which has contributions from electrons and phonons. It is clear from $ZT$ definition that in order to improve thermoelectric figure of merit one should increase the thermopower $S^2\sigma$ and decrease the thermal conductivity. Different approaches have been tried in order to enhance the thermoelectric properties of $Bi_2Te_3$ or its alloys. These approaches included the composition change from its stoichiometry, the use of polycrystalline materials with different grain sizes, intentional introduction of structural defects and incorporation of different dopants, e.g. Sb or Se, into $Bi_2Te_3$ lattice. The optimization of bulk $Bi_2Te_3$ led to incremental improvements but no breakthrough enhancement in $ZT$ was achieved.

More promising results ($ZT$ ~2.4 for p-type material at RT) were achieved with $Bi_2Te_3/Sb_2Te_3$ superlattices produced by the low-temperature deposition [4]. A recent study indicated that the low-dimensional structuring of BiSbTe alloys [5] also allows for $ZT$ enhancement to ~1.5 at RT. But still higher $ZT$ values are needed for a major practical impact. It has been shown that $ZT$ above 3 or 4 at RT are needed in order to make thermoelectric cooling or power generation competitive with conventional methods [6]. Such an increase in $ZT$ would lead to a "thermoelectric revolution" and allow one for much more environmentally friendly power generation and cooling.

It follows from many theoretical predictions that a drastic improvement in $ZT$ can be achieved in low-dimensional structures where electrons (holes) are strongly confinement in one or two dimensions [7]. Hicks and Dresselhaus [8] predicted that $ZT$ can be increased in $Bi_2Te_3$ quantum well by a factor of ~13 over the bulk value. This would require a complete carrier confinement in a quantum well with a width $H$ on the order of ~1 nm and an optimized position of the Fermi level. According to Dresselhaus *et al*. [7-8], quantum confinement of charge carries in quantum wells leads to a drastic *ZT* improvement due to the increase in the carrier density-of-states (DOS) near the Fermi level and corresponding increase in the thermopower. The crucial condition for such mechanism is quantum confinement of carriers in quantum wells, which is only possible if materials are crystalline and essentially free of defects. The thickness of the thin film required to achieve the quantum confinement conditions has to be on the order of few atomic layers. Note that in the superlattices commonly used in thermoelectric studies the charge carries are only partially confined if confined at all due to the small potential barrier height and relatively low material quality. The barrier height has a pronounced effect on *ZT*. Broido and Reinecke [9] have shown theoretically that *ZT*=3 can be achieved in $Bi_2Te_3$ superlattices with infinite potentials when the quantum well width (i.e. thickness of the thin film) is *H*~3 nm. In the structures with incomplete quantum confinement the maximum *ZT* decreases to ~2.5 and the required width becomes as small as ~2 nm.

Balandin and Wang [10] proposed a different strategy for increasing *ZT* in low-dimensional structures by reducing its thermal conductivity via spatial confinement of acoustic phonons, which carry bulk of heat in thermoelectric materials. The improvement of thermoelectric properties via phonon engineering [10-11] also can be achieved in thin films or nanowires with the thickness of just few atomic layers and high quality of interfaces. In nanostructures with rough interfaces, the thermal conductivity can be reduced due to phonon scattering on boundaries and defects [12-14]. At the same time, defects and disorder can also lead to electron mobility degradation limiting the improvement. The phonon – confinement mechanism of the thermal conductivity reduction, proposed by Balandin and Wang [10], originates from the decreased phonon group velocity of the confined acoustic phonon modes, which results in the increased scattering on point defects [10-11]. This mechanism works even in atomic films with smooth interfaces and can be utilized without strongly degrading the electron mobility.

Thus, in order to employ the full strength of the low-dimensional confinement effects for improving thermoelectric figure of merit either via the electron band-structure and phonon engineering one needs to produce quasi-two-dimensional (2D) structures with a few-atomic layer thickness and high quality interfaces. Conventional chemical vapor deposition, electrochemical or other means are not capable for producing such quality structures. Molecular beam epitaxy (MBE) of low-dimensional thermoelectric materials was also much less successful than that of optoelectronic or electronic materials due to the lattice mismatch and other factors. These considerations create very strong motivations for the search of alternative approaches to fabrication of the stacks of quasi-2D crystals made of $Bi_3Te_3$-based materials.

Few years ago an interest to the stacked quasi-2D layers of bismuth telluride received an additional impetus from a totally different direction. It has been shown that stacks of quasi-2D layers of Te-Bi-Te-Bi-Te are members of a new type of recently discovered materials referred to as *topological insulators* [15-16]. The surface state of a quasi-2D crystal of $Bi_2Te_3$ is predicted to consist of a single Dirac cone. Moreover, it has been shown that the layered structures of related materials such as $Bi_2Se_3$ and $Sb_2Te_3$ are also topological insulators. The particles in topological insulators coated with thin ferromagnetic layers have manifested exotic physics and were proposed for possible applications in the magnetic memory where write and read operations are achieved by purely electric means. All these stimulate the search for methods to produce quasi-2D crystals of bismuth telluride even further. The mechanically exfoliated atomically-thin films of bismuth telluride can be transferred to various substrates and coated with other materials. Most recently, a brief report, we have shown a possibility of cleavage from bulk $Bi_2Te_3$ of thin films and ribbons with the thickness of few-atomic layers [17]. The electrical measurements revealed a rather unusual temperature dependence of their current-voltage characteristics.

In this letter, we show that individual atomically-thin layers of bismuth telluride can be mechanically exfoliated from bulk $Bi_2Te_3$ in a "*graphene-like*" fashion. The presence of the van der Waals gaps allowed us to disassemble $Bi_2Te_3$ crystal into its *quintuple* building blocks – five mono-atomic sheets of $Te^{(1)}$-Bi-$Te^{(2)}$-Bi-$Te^{(1)}$, which have the thickness of ~1 nm. Moreover, our microscopy data indicate that the intra-quintuple bonds can be broken further leading to quasi-2D crystals. To the best of our knowledge, this is the very first successful exfoliation of the large-area few-atom-thick layers of crystalline bismuth telluride or related material. The resulting quasi-2D crystals retain their good electrical conduction and poor thermal conduction properties important for thermoelectric applications. Considering that $Bi_2Te_3$ materials for the thermoelectric industry are as important as silicon (Si) and silicon oxide ($SiO_2$) for the electronic industry our results may have significant practical impact. The exfoliation procedure can become a new enabling technology for producing stacks of the low-dimensional thermoelectric films with complete quantum confinement of charge carriers and acoustic phonons. The developed technology for producing free-standing quasi-2D layers of $Te^{(1)}$-Bi-$Te^{(2)}$-Bi-$Te^{(1)}$ creates an impetus for investigation of topological insulators and their possible practical applications.

$Bi_2Te_3$ has the rhombohedral crystal structure of the space group $D_{3d}^5$ – R(-3)m with five atoms in one unit cell. The lattice parameters of the hexagonal cells of $Bi_2Te_3$ are $a_H$= 0.4384 nm and $c_H$ = 3.045 nm [14]. Its atomic arrangement can be visualized in terms of the layered sandwich structure (see Figure 1). Each sandwich is built up by five mono-atomic sheets, referred to as *quintuple* layers, along the $c_H$ axis with the sequence – [$Te^{(1)}$-Bi-$Te^{(2)}$-Bi-$Te^{(1)}$] – [$Te^{(1)}$-Bi-$Te^{(2)}$-Bi-$Te^{(1)}$] –. Here superscripts (1) and (2) denote two different chemical states for the anions. The outmost atoms $Te^{(1)}$ are strongly bound to three planar $Te^{(1)}$ and three Bi metal atoms of the same quintuple layers and weakly bound to three $Te^{(1)}$ atoms of the next sandwich.

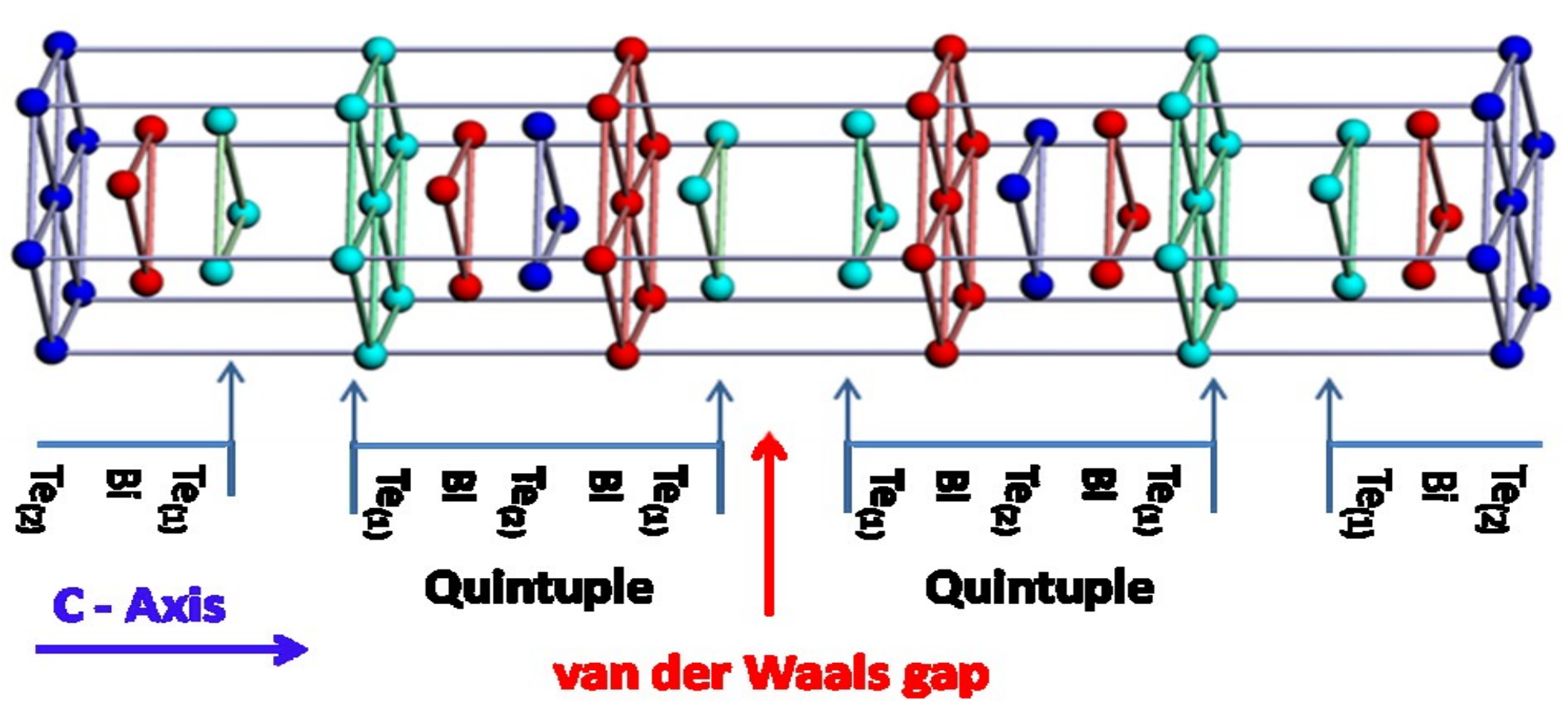

**Figure 1:** Schematic of $Bi_2Te_3$ crystal structure of $D_{3d}^5$ – R(-3)m space group showing quintuple layers and location of the van der Waals gaps. The $Te^{(1)}$-$Te^{(1)}$ bond is the weakest while Bi-$Te^{(1)}$ bond is the strongest. The mechanical exfoliation mostly results in breaking the $Te^{(1)}$-$Te^{(1)}$ van der Waals bond and formation of quintuples although in some cases intra-quintuples bonds also break leading to bi-atomic and tri-atomic layers.

The binding between adjacent $Te^{(1)}$ layers originates mostly from the weak van der Waals forces although other long-range Coulomb forces play role in the bonding. The stronger bonds inside the quintuple layers are of the covalent or partially ionic nature. The presence of the van der Waals gap between the quintuples results in easy cleavage plane between the adjacent $Te^{(1)}$ layers. The bond strength within the quintuple layers is not the same. The Bi-$Te^{(1)}$ bond is stronger than Bi-$Te^{(2)}$ bond, which is the second weakest points within the crystal structure. It is believed that the Bi-Te(2) bond is covalent while the Bi-Te(1) binding includes both covalent and ionic interaction. The lattice spacing between layers has a direct relationship with the atomic bond strength between the neighboring layers. For this reason the weakest $Te^{(1)}$-$Te^{(1)}$ bond correspond to the largest spacing *d*~0.37 nm. What is also important for our purposes is that the strength and length of Bi-$Te^{(2)}$ bond is not much different from the van der Waals gaps of $Te^{(1)}$-$Te^{(1)}$. The latter suggests that the mechanical exfoliation may lead not only to [$Te^{(1)}$-Bi-$Te^{(2)}$-Bi-$Te^{(1)}$] quintuples but also to separate atomic planes of Bi-Te and Te-Bi-Te. One should note here that for thermoelectric applications, the quintuple layers are of greater interest than single atomic planes of Bi or Te atoms. For this reason, in this study are mostly interested in producing individual quintuples or few-quintuple layers. The quasi-2D quintuple layers are also of principle importance for the investigation of topological insulators.

In order to isolate bismuth telluride quintuples and break them into atomic planes we employ a method similar to the one used for exfoliation of single-layer graphene [18-21]. Through a mechanical cleavage process we separated thin films from crystalline bulk $Bi_2Te_3$. The process was repeated several times to obtain the layers with just few-atomic planes. Owing to the specific structure of $Bi_2Te_3$ crystal along $c_H$ direction ($c_H$ = 3.045 nm is very large lattice constant as compared to other materials) we were able to verify the number of layers using the optical inspection combined with the atomic force microscopy (AFM) and the scanning electron microscopy (SEM). The thickness of the atomic quintuple is *H*~1 nm. The step-like changes in the cleaved layers highest of ~1 nm can be distinguished well with AFM. The produced atomic layers were placed on Si substrates containing 300-nm thick $SiO_2$ capping layer. The silicon oxide thickness was selected by analogy with that for graphene on Si/$SiO_2$ [18-21]. We then isolated and separate individual crystal planes which exhibited high crystal quality with little to no structural defects. The mechanical cleavage of $Bi_2Te_3$ has led to a portion of quintuples with five atomic planes. In some cases we observed atomic planes with the thickness smaller than that of the quintuple layers. The produced flakes had various shapes and sizes ranging from ~2 μm to 30 μm. Some flakes had correct geometrical shapes indicative of the facets and suggesting the high degree of crystallinity. We selected large uniform Bi-Te flakes with the dimensions of ~ 20 – 30 μm for fabrication of metal contacts for electrical characterization. For detail material characterization, we suspended some of the Bi-Te atomic films over trenches in $SiO_2$/Si substrates. The trenches were fabricated by the reactive ion etching (RIE). They had a depth of ~300 nm and widths ranging from 1 to 5 μm. By suspending the ultra-thin atomic films over these trenches we reduced the coupling to the substrate. The latter allowed us to achieve better understanding of the intrinsic properties of the atomically thin layers.

The isolated Bi-Te atomic layers were investigated using a high-resolution field-emission scanning electron microscope (XL-30 FEG) operated at 10-15 kV. The diffraction patterns of the crystalline structures of the layers were studied using transmission electron microscopy (TEM). The sample preparation for TEM (FEI-PHILIPS CM300) inspection was carried out through ultrasonic separation in ethanol ($C_2H_5OH$) solution. The sonication was most effective with 500 μL of $C_2H_5OH$ solution where the molar concentration of Bi-Te films was held at a constant 1.41 x $10^{-1}$ moles/liter throughout the solution. AFM studies were performed using VEECO instrument with the vertical resolution down to ~0.1 nm. Raman spectroscopy was performed for better material identification and characterization of intrinsic properties of the resulting films. A micro-Raman spectrometer (Renishaw RM 2000) was used in a backscattering configuration excited with a visible laser light (λ = 488 nm). The spectra were collected through a 50X objective and recorded with 1800 lines/mm grating providing the spectral resolution of ~ 1 $cm^{-1}$ (software enhanced resolution is ~0.5 $cm^{-1}$). All spectra were recorded at very low power levels $P$<0.5 mW measured with the power meter (Orion) at the sample surface. The low power levels were essential to avoid local laser heating, which was much more pronounced than in other material systems due to the extremely low thermal conductivity of $Bi_2Te_3$. The details of our Raman instrumentation and measurement procedures were reported by us earlier in the context of graphene investigation [22-24]. The electrodes and contact pads for the ultrathin Bi-Te layers on $SiO_2$ were fabricated and defined by the electron beam lithography (EBL) system (Leo 1550) followed by the metal deposition of 7-nm/70-nm Ti/Au metals with the electron beam evaporator (Temescal BJD-1800). The electrical measurements were carried out using a probe station (Signatone) at ambient conditions.

The quintuples, atomic tri-layers and bi-layers we were identified via combined optical, SEM, AFM and TEM inspection (see Supplemental Materials). Our prior extensive experience with graphene and few-layer graphene [22-24] was instrumental in our ability to separate weakly distinguishable optical features in Bi-Te atomic layers placed on $SiO_2$/Si substrates. A relatively large thickness of the unit cell ($c_H$ = 3.045 nm) made SEM and AFM identification rather effective. The representative high-resolution SEM micrographs and AFM images presented in Figure 2 (a-f) show SEM, AFM and TEM images of Bi-Te quasi-2-D crystals with the lateral sizes ranging from a few microns to tens of microns. In Figure 2 (a), one can see SEM micrograph of high quality crystalline films with the lateral dimensions of 1-4 μm and a thickness of few-atomic planes. Due to the atomic thickness of the films the overlapping regions are seen with a very high contract. The larger area flakes (>20 μm) tended to be attached to thicker $Bi_2Te_3$ films (Figure 2 (b)). It was our experience that for thicker films the use of the tilted SEM images was more effective in visualizing the quintuples and atomic-layer steps. Although [$Te^{(1)}$-Bi-$Te^{(2)}$-Bi-$Te^{(1)}$] quintuple layers were more readily available among the separated flakes, some regions of flakes had sub-1-nm thickness and appeared as protruding few-atomic-plane films. Figures 2 (c-d) show suspended few-quintuple films. In Figure 2 (e) we present a typical AFM image of steps in the separated films with clear layered structure. One step with the height of ~1 nm corresponds to a quintuple while two steps per a profile height change of ~ 1 nm indicate sub-quintuple units (e.g. bi-layer and a tri-layer). The structural analysis at the nanometer scale was carried out using TEM and the selected area electron diffraction (SAED) technique. The images for this type of characterization were taken with the electron beam energy of 300 kV. The exfoliated samples were dissolved in ethanol and placed on copper grids. The TEM image in Figure 2 (f) shows that the exfoliated Bi-Te flakes dissolved in solution retain their flat structure and do not form clusters after being subjected to ultrasonic vibrations and processing.

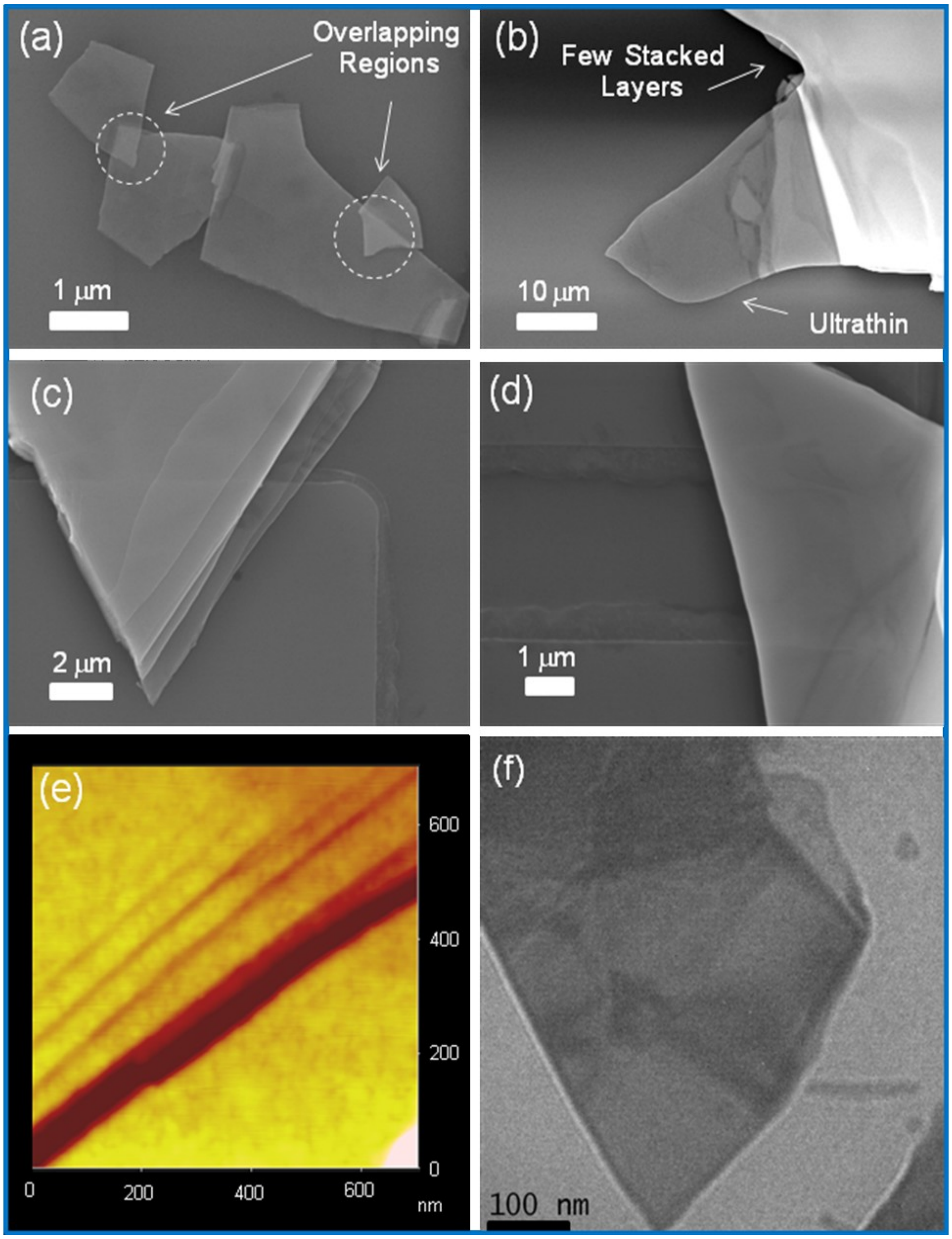

**Figure 2**: Images of quasi-2D bismuth telluride crystals showing (a) SEM micrograph of the overlapping few-layer Bi-Te atomic crystals; (b) large-area atomically-thin crystal attached to thick $Bi_2Te_3$ film; (c) suspended films with visible quasi-2D layers; (d) suspended few-atom-plane film; (e) AFM micrograph of few-atomic-layer steps in the cleaved films; and (f) TEM micrograph of the quasi-2D bismuth telluride film.

In Figure 3 (a) we present an electron diffraction pattern, which indicates that the separated atomically-thin layers of bismuth telluride are crystalline after all processing steps. The elemental composition and stoichiometry of the atomically thin Bi-Te layers were studied by the energy dispersive spectrometry (EDS). The data were recorded for the suspended films (like those shown in Figure 2 (c-d)). Representative EDS spectra for a suspended quasi-2-D crystal and a thick $Bi_2Te_3$ film (used as a "bulk" reference) are shown in Figure 3 (b-c). Note that the most pronounced peaks in Figure 3 (b) are those of Si and oxygen (O) proving the electron beam penetration through the suspended atomically-thin film. The Si and O peaks are absent in Figure 3 (c) for thick $Bi_2Te_3$ crystal, which absorbes the elecrton beam completely.

For one of the films, which had a thickness of ~40 atomic planes (~8 quintuple layers), and occupied an intermediate position between quasi-2D crystals and bulk $Bi_2Te_3$ crystals we found that the molar contents of Bi and Te were found to be ~34.7% and 65.3%, respectively. Although the film is crystalline and the stoichiometry for the [$Te^{(1)}$-Bi-$Te^{(2)}$-Bi-$Te^{(1)}$] quintuple layers is preserved, one can talk about an apparent deviation from the stoichiometry due to the fact that the film is not truly bulk. For the large bulk bismuth telluride crystals used for exfoliation of the ultra-thin films we consistently found 40% to 60% ratio of Bi to Te corresponding to $Bi_2Te_3$ formula. The analysis of the measured EDS spectrum of the suspended film, which made up ~10% of the total wt%, had a molar content percentage between Bi and Te of ~33.2% and 66.8%, respectively. Thus, the structural make up of that particular ultra-thin film has 1 to 2 ratio of Bi to Te atoms (e.g. Te-Bi-Te), which differs strongly from conventional bulk crystals.

The possibility of changing the "effective" atomic composition in the crystalline ultra-thin films is very important for practical applications. It is known from the extensive studies of the thermoelectric properties of bismuth telluride, that a small small variation of ~0.1% in the Bi to Te atomic ratio can change the properties of the material from p-type to n-type [25]. Intentional deviation from stoichiometry during synthesis of bismuth telluride compounds and alloys has been conventionally used for "doping" this type of materials [26]. The close-to-stoichiometric $Bi_2Te_3$ is of p-type with a free carrier concentration of approximately $10^{-19}$ $cm^{-3}$. A shift to excess Te leads to an n-type material. Since our atomically thin films are crystalline, the "stoichiometric doping" may work in different ways than in the polycrystalline or disordered alloy bismuth telluride materials. In the atomically-thin crystals the charge can accumulate on the film surfaces or film – substrate interfaces. In this sense, the obtained quasi-2D crystals may open up a new way for doping and tuning the properties of bismuth telluride materials.

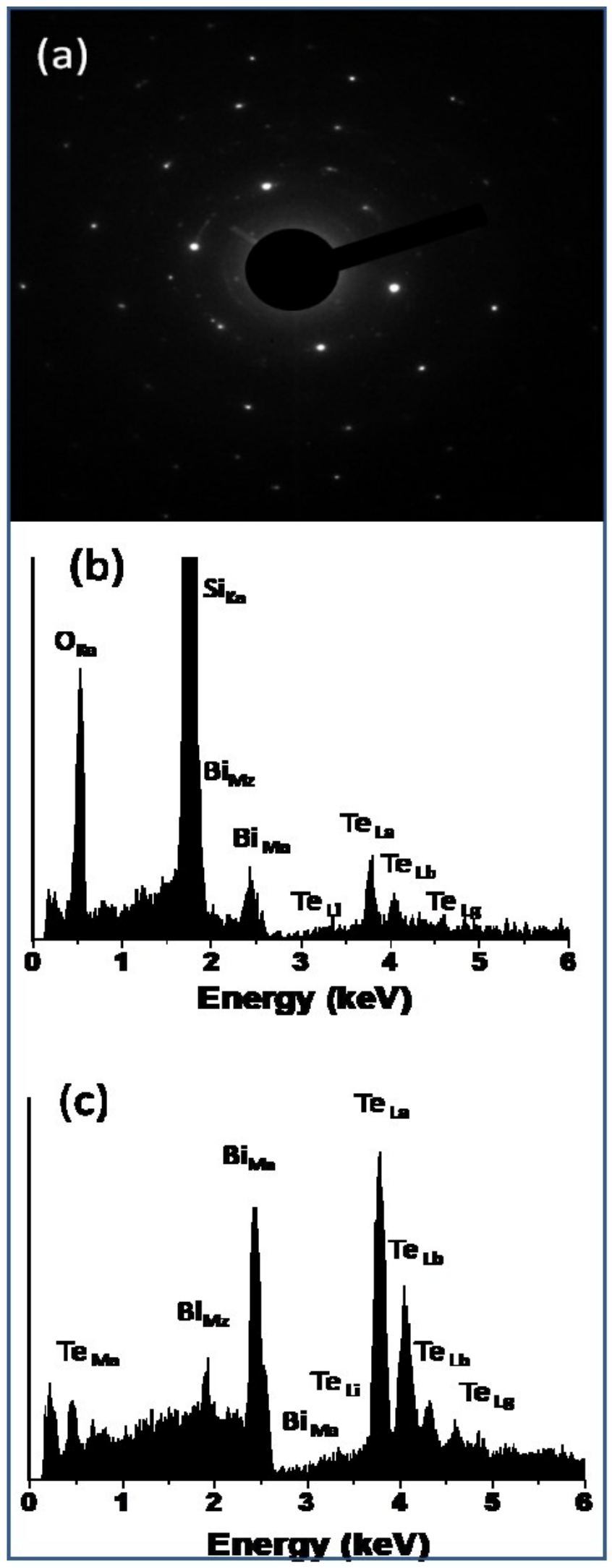

**Figure 3**: Structural and compositional characterization data showing (a) electron diffraction pattern indicating that quasi-2D Bi-Te films are crystalline; (b) EDS spectrum of the suspended atomic film of bismuth telluride; and (c) EDS spectrum of the reference thick film. Note that the dominant peaks in the EDS spectrum of the quasi-2D Bi-Te film shown in (b) are those of Si and O proving transparency of the atomic film for the electron beam. These peaks are absent in the spectrum of the reference bulk $Bi_2Te_3$.

We now turn to the analysis of Raman spectrum of the atomically-thin bismuth telluride films. $Bi_2Te_3$ has five atoms in its unit cell and, correspondingly, 15 phonon (lattice vibration) branches near the Brillouin zone (BZ) center (phonon wave vector $q$=0) [27]. Twelve of those 15 branches are optical phonons while the remaining 3 are acoustic phonons with the $A_{2u}$ + $E_u$ symmetry. The 12 optical modes are characterized by $2A_{2u} + 2E_u + 2A_{1g} + 2E_g$ symmetry. Each of the $E_g$ and $A_{1g}$ modes are two-fold degenerate. In these phonon modes, the atoms vibrate in-plane and out-of-plane (i.e. perpendicular to the film plane), respectively [28]. We focused our analysis on the phonon peaks in the spectral range from 25 $cm^{-1}$ to 250 $cm^{-1}$. Figure 4 (a-b) show SEM image of the Bi-Te atomic film with the locations where the Raman spectra were taken as well as the spectra themselves. These spectra were recorded at very low excitation laser power (~0.22 mW on the sample surface) to avoid local heating. The examined flake was placed in such a way that it had a suspended region as well as regions rested on Si and $SiO_2$. The observed four Raman optical phonon peaks were identified as $A_{1g}$ at ~ 62 $cm^{-1}$, $E_{g2}$ at ~104 $cm^{-1}$, $A_{1u}$ at ~120 $cm^{-1}$ and $A_{2g}$ at ~137 $cm^{-1}$. These peak positions are very close to the previously measured and assigned Raman peaks of bulk crystalline $Bi_2Te_3$ [27-29]. Richter *et al.* [27] in their detail study of phonons in $Bi_2Te_3$ provided the following frequencies (in their notation): 134 $cm^{-1}$ for A, 103 $cm^{-1}$ for $E_g$ and 120 $cm^{-1}$ for $A_{1u}$. The $A_{1u}$ peak is likely to become Raman active due to the symmetry breaking in atomically thin films. One can notice that the out-of-plane vibrations (at ~137 $cm^{-1}$ and ~119 $cm^{-1}$) in the suspended Bi-Te atomic films have higher intensity. The latter may be an indication of the enhancement of these vibration modes in the atomically thin films, which are not supported by the substrate. The Raman study confirms that the exfoliated films are crystalline and atomically thin. A systematic study of the changes in Raman spectrum due to modification of the vibration modes in the exfoliated ultra-thin films suspended or supported on the substrate was complicated due to pronounced local heating effects.

Bulk single crystal $Bi_2Te_3$ is known to have very low thermal conductivity of ~ 1.5 – 2.0 W/mK along the cleavage plane and 0.6 W/mK along the van der Waals bonding direction [30]. It also has a low melting point of 573°C. The local laser heating was a major problem when we tried to increase the excitation power to the levels conventionally used for other materials.

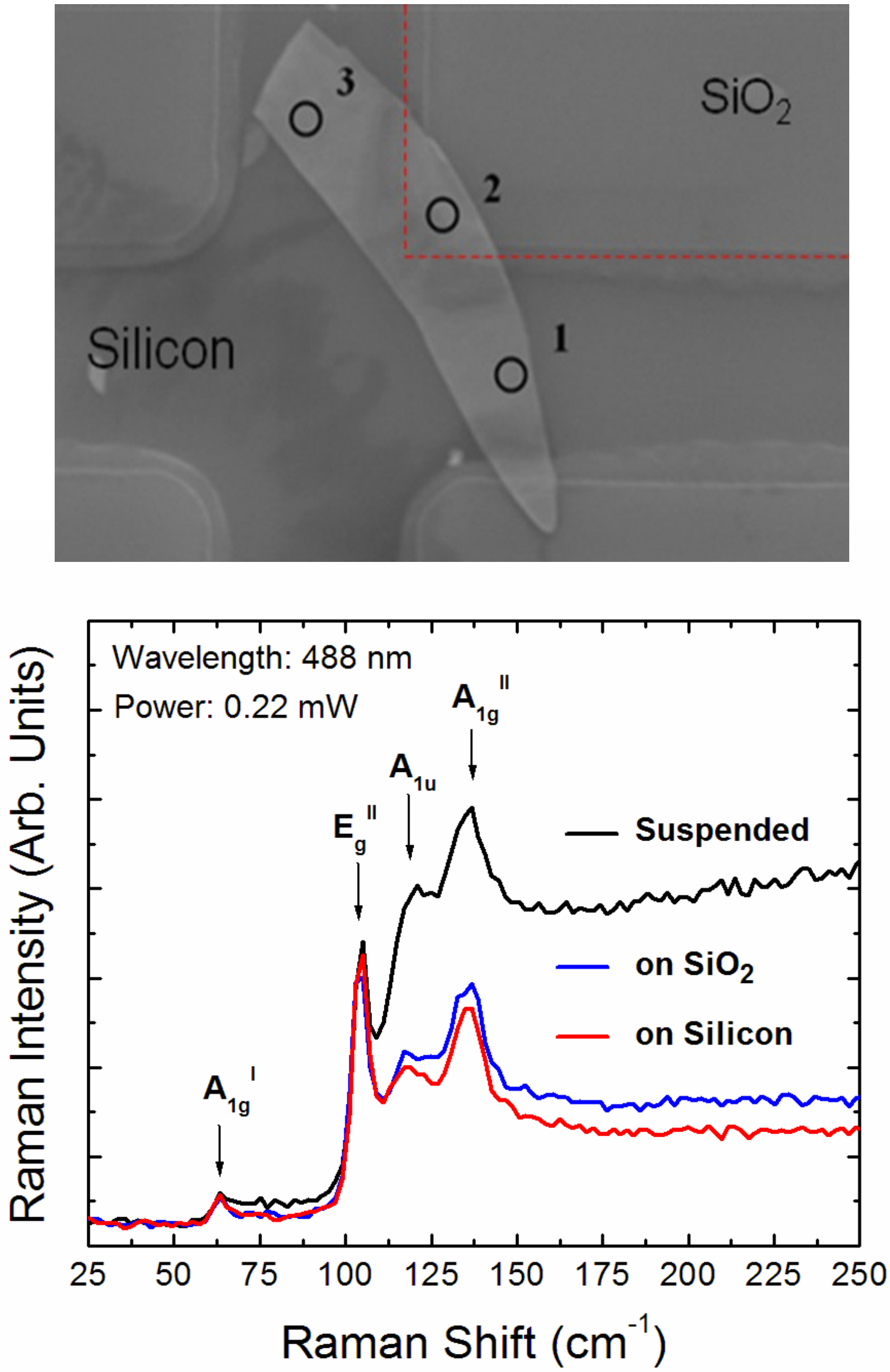

**Figure 4**: Raman spectra of quasi-2D bismuth telluride crystals. (a) SEM image showing suspended Bi-Te atomic film, which rests partially on $SiO_2$ and Si regions of the substrate. The spectra from the suspended and supported regions were recorded in the locations marked as 1, 2, and 3. (b) Informative Raman bands in the spectra of Bi-Te atomic films recorded at very low excitation power level. Note that the out-of-plane phonon modes in the suspended atomic crystals have higher intensity.

The maximum excitation power of the Ar+ laser (with the wavelength of 488 nm) used in this study was 10 mW. Approximately half of the excitation power reaches the sample surface after transmission through the optical system. The use of the power levels above 0.5 mW (corresponded ~0.22 mW on the sample) resulted in appearance of the holes due to local melting or oxidation of the atomically-thin flakes as it is seen in the inset to Figure 5. In this image 100% corresponds to the power of 10 mW set at the laser. We observed reproducibly that the diameter of the laser burned holes was growing with increasing excitation power (the exposure time for each spot was 100 seconds).

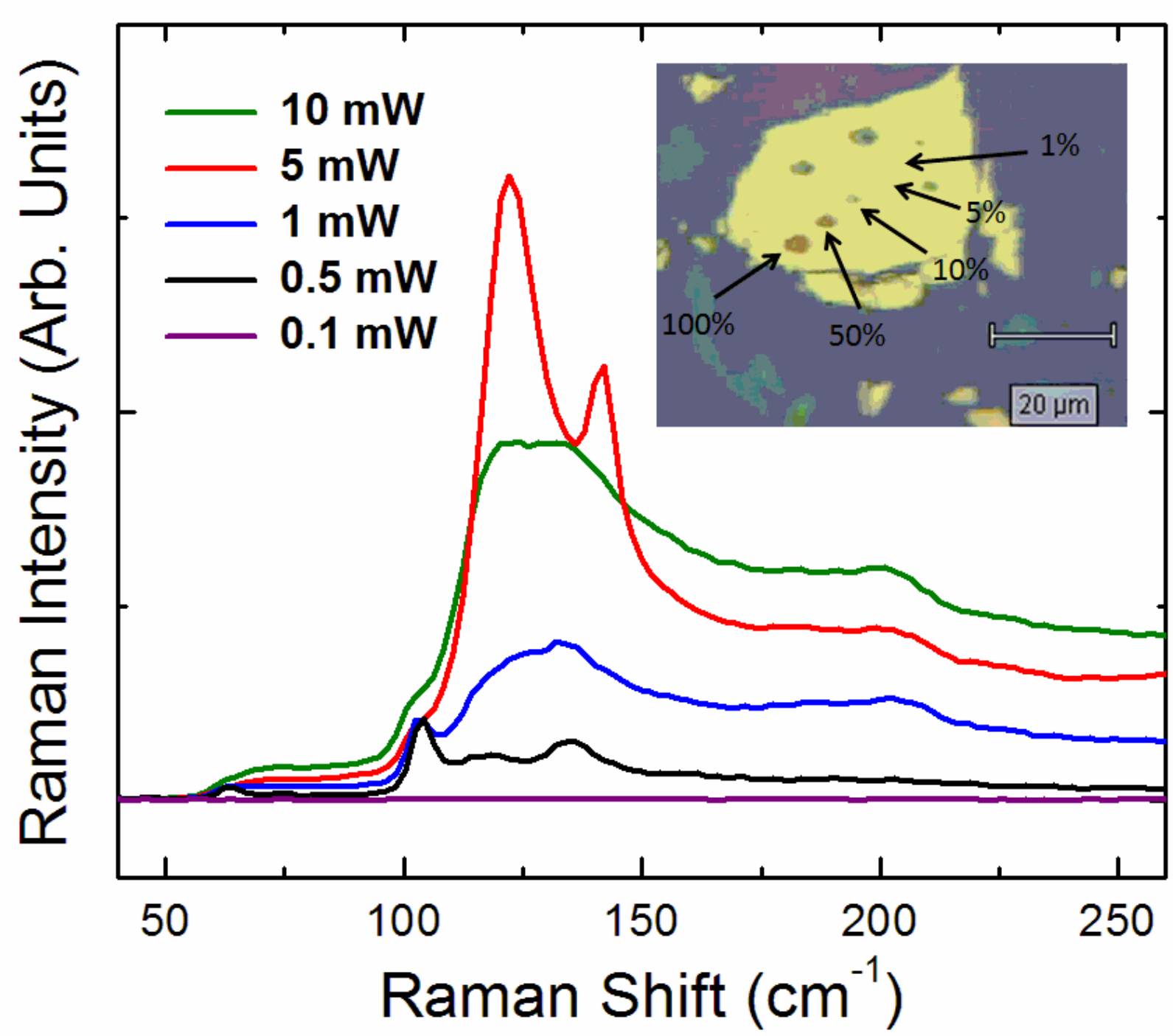

**Figure 5**: Evolution of Raman spectra from the Bi-Te atomic film with changing intensity of the excitation laser power illustrating a very narrow range of power levels suitable for exciting informative phonon bands. The power levels above 0.5 mW (corresponding to ~ 0.22 mW on the sample surface) lead to local melting of the atomic Bi-Te crystals in sharp contrast to graphene. The inset shows the spots from which the Raman spectra were recorded. Note the burned holes in the atomic films when the excitation power is above 0.5 mW.

Figure 5 shows an evolution in Raman spectra of Bi-Te atomic layers as the excitation power changes. When the power is too low (0.1 mW) no spectrum is excited. The excitation power of 0.5 mW provides meaningful spectra, which are in line with those obtained for bulk $Bi_2Te_3$ (at this power level no hole-burning or other laser-induced structural disorder were observed in microscopy images). As the power grows higher than 0.5 mW the Raman spectrum begins to change as a result of the local melting of the material. No laser damage was observed for the bulk $Bi_2Te_3$ at these power levels. For this reason, the selection of the right excitation power for Raman spectroscopy of the atomically thin Bi-Te layers is crucial for obtaining informative phonon bands. The drastically different reaction of the films on laser heating confirms their few-atomic layer thickness. In principle, one can envision a method for verifying the thickness of the film by examining the power dependence of the diameter of the burned holes.

The observed easy local heating damage to Bi-Te atomic flakes was in sharp contract to graphene, a material characterized by extremely high thermal conductivity [31-32]. We were able to use much higher laser power on suspended graphene without inflicting any damage to its lattice. The heat conduction in strictly 2D systems is a complicated subject deserving special consideration. The thermal conductivity of conventional thin films usually decreases with decreasing film thickness as a result of the acoustic phonon – boundary scattering [12-13]. At the same time, it is also known that the thermal conductivity limited by crystal anharmonisity (also referred to as intrinsic) has a logarithmic divergence in strictly 2D system [33]. This anomalous behavior of the 2D thermal conductivity has been studied extensively for many different crystal lattices and atomic potentials [34-35]. One needs disorder (e.g. extrinsic scattering mechanisms) in order to obtain a finite value of the thermal conductivity in 2-D systems or limit the lateral size of the system [33-35]. In our case of Bi-Te flakes, the samples had the thickness of several atomic layers (not exactly 2D system but rather a quasi-2D) and the extrinsic effects were dominant. Due to the very low thermal diffusivity and thermal conductivity of $Bi_2Te_3$ crystals at first place, the induced heat had not escaped fast enough from the local spots leading to the lattice melting and observed lattice damages. As a consequence, for thermoelectric applications, it would be better to use stacks of bismuth telluride quintuple layers, put one on top of the other, rather than atomic bi-layers. Indeed, quintuples are more mechanically robust and expected have even lower thermal conductivity than $Bi_2Te_3$ bulk values.

It is important to understand if the electrical conductivity of bismuth telluride is preserved after it was structured to films with the thickness of just few atomic layers. For transport measurements were prepared Bi-Te devices with a bottom gate and two top metal contacts (see inset to Figure 6). The RT current – voltage characteristics shown in Figure 6 reveal a weak non-linearity and

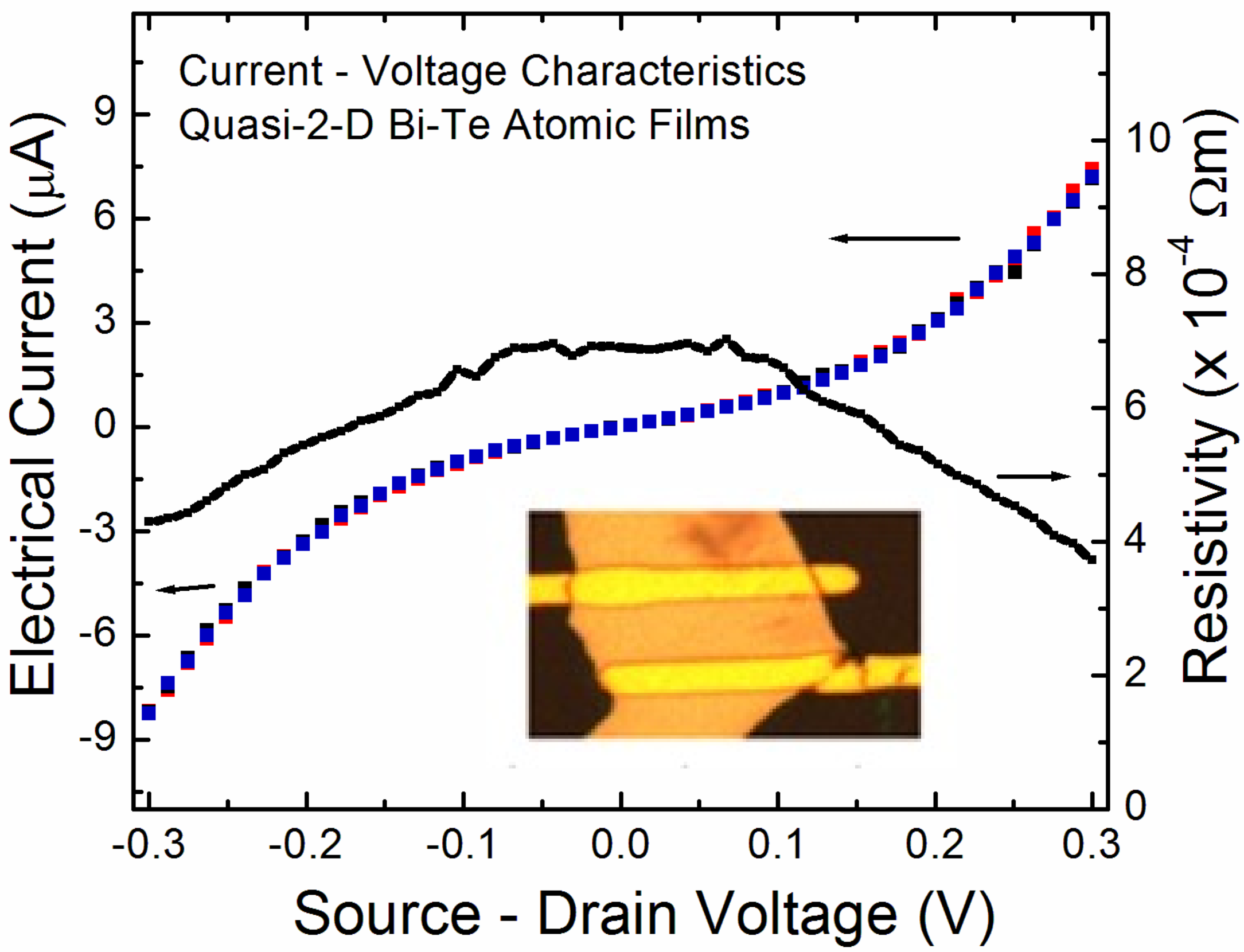

**Figure 6:** Electrical current and resistivity of the quasi-2D bismuth telluride crystal as functions of the applied source – drain bias. The inset shows an optical microscopy image of the test structure.

This value (which includes contact resistance) is comparable to the resistivity values frequently measured for thick evaporated $Bi_2Te_3$ films used in thermoelectric devices [36-37]. The latter suggests that the charge carriers are not depleted in the samples and that the atomically thin bismuth telluride films retain their electrical properties. We did not observe a strong gating effect for Bi-Te devices while sweeping the back gate bias fro -50V to 50V. That was in a sharp contract to our experiments with the back-gated graphene and few-layer graphene devices [38-39]. In fabrication of graphene and Bi-Te devices we followed similar procedures and used the same heavily-doped $Si/SiO_2$ wafers. One possible explanation of the weak gating in Bi-Te atomic films can be a strong doping of the flakes due to the "stoichiometric doping" discussed above. The charge accumulation at the interfaces terminated with either Bi or Te atoms can screen the electric field produced by the back gate.

The described electrically conducting bismuth telluride quasi-2D crystals can be used as quantum wells with nearly infinite potential barriers for thermoelectric applications. The charge carriers and acoustic phonons in the crystalline quintuple layers with the thickness of ~1 nm will be strongly confined spatially. At the same time, any practical application of thermoelectric nanostructures requires a sufficient "bulk" volume of the material for development of the temperature gradient, in case of cooling, or voltage, in case of power generation. An individual quantum well would not be suitable. For this reason, we envisioned a practical method for utilization of Bi-Te quasi-2D crystals by stacking exfoliated films one on top of the other [40]. The obtained "*stacked superlattices*" are expected to retain the useful properties of individual atomic films such as quantum confinement of charge carriers and reduced phonon thermal conductivity. Indeed, the potential barriers for charge carries in such crystalline films remain very high unlike in the epitaxially grown $Bi_2Te_3$-based superlattices with the lattice matched barriers where the band off-sets are small. The thermal conductivity in the mechanically separated and stacked layers is reduced due to the acoustic phonon – rough boundary scattering or acoustic phonon spatial confinement in exactly the same way as in the individual films. We experimentally tested a possibility of *ZT* enhancement in prototype stacked films by measuring the thermal conductivity and Seebeck coefficient, and comparing them with those of the reference $Bi_2Te_3$ bulk crystals, which was used for the exfoliation of the films.

The thermal conductivity *K* of the prototype stacked atomic-films was determined by two different methods: the laser-flash (Netszch LFA) and the transient planar source (TPS) technique (Hot Disk). Both techniques have been "calibrated" via comparison with the values obtained with the in-house built 3-ω method [41-43], which is considered to be a standard technique for thin films. We have previously successfully used 3-ω measurements for electrically insulating thin films with the thickness down to ~1 nm [44]. The Seebeck coefficient of stacked films was determined using MMR system (SB100) consisting of two pairs of thermocouples. Details of the measurements are provided in the Methods. We found significant drop in the cross-plane (in-plane) thermal conductivity from ~0.5 – 0.6 W/mK (~1.5 – 2.0 W/mK) in bulk reference to ~0.1-0.3 W/mK (~1.1 W/mK) in the stacked films at RT. The thermal conductivity of stacked superlattices revealed a very weak temperature dependence suggesting that the acoustic phonon transport was dominated by the boundary scattering. It is interesting to note that the measured cross-plane *K* value is close to the minimum thermal conductivity for $Bi_2Te_3$ predicted by the Cahill model [45], which gives the low bound of ~0.12 W/mK. We have not observed substantial changes in the thermal power beyond experimental uncertainty with the measured Seebeck coefficient (~234 μV/K) only slightly above its bulk reference value. The latter was attributed to the fact that the Fermi level was not optimized in these films. More research is needed to gain complete control of the carrier densities in the mechanically exfoliated films. At the same time, the measured decrease of the thermal conductivity results in the ZT enhancement by about 30–40 % of its bulk value. Additional *ZT* increase in a wide temperature range can be achieved with the cross-plane electrical gating of the Bi-Te atomic films. Some of us have recently shown theoretically [46] that a combination of quantum confinement of carriers and perpendicularly applied electric field in bismuth telluride nanostructures can be effective for ZT improvement. The developed exfoliation technique can also be extended to other thermoelectric material systems [47]. This approach is particularly promising for the thermoelectric cooling applications at low temperature because the effect is even more pronounced owing to stronger suppression of the thermal conductivity.

In conclusion, following the procedures similar to those developed for graphene mechanical exfoliation, we were able to produce bismuth telluride crystals with a thickness of few atoms. The quais-2D atomic crystals were suspended across trenches in $Si/SiO_2$ substrates and subjected to detail characterization using SEM, TEM, EDS, AFM, SAED and micro-Raman spectroscopy. It was established that the presence of the van der Waals gaps in $Bi_2Te_3$ crystals allows one to disassemble them into atomic quintuples, i.e. five atomic planes Te-Bi-Te-Bi-Te, which build up 3D crystal. Moreover the microscopy analysis shows that the bonds inside quintuples can be broken further leading to Bi-Te bi-layers and Te-Bi-Te tri-layers. By altering the thickness and sequence of atomic planes one can create "designer" non-stoichiometric crystalline films and change their properties. The exfoliated quintuples have low thermal conductivity and good electrical conductivity. The "stacked superlattices" made of the mechanically exfoliated bismuth telluride films show enhanced thermoelectric properties. The obtained results may lead to completely new scalable methods for producing low-dimensional thermoelectrics and atomic-layer engineering of their properties. The described technology for producing free-standing quasi-2D layers of Te-Bi-Te-Bi-Te can be used for investigation of the topological insulators and their possible practical applications.

**Methods**

The first technique used for $K$ measurements is the laser-flash technique (Netszch LFA). By timing the heat pulse propagation through our samples of known thicknesses we applied the transient method to measure the thermal diffusivity and thermal conductivity. The experimental setup was equipped with an adjustable Xenon-Flash-Lamp for heating the sample on one end while a contactless IR detector was used to measure the temperature rise at the other end. The specific instrument used for this study had a capability to measure $K$ in the range from below 0.1 W/mK to 2000 W/mK. We have previously "calibrated" the laser-flash measurement technique with other standard methods for measuring thermal conductivity available in our laboratory [41-44]. Due to the geometry of the setup the measured $K$ has to be interpreted as mostly cross-plane component of the thermal conductivity tensor. In the experiments the light pulse "shot" hit the sample and a temperature rise (up to 2 K) was measured by an InSb IR detector. The thermal conductivity is defined as $K=\alpha\rho C_p$, where $\alpha$ is the thermal diffusivity of the film determined in the experiment as $\alpha = 0.139\times Z^2/t_{1/2}$, $t_{1/2}$ is the measured half-rise time of temperature, $C_p$ is the heat capacity, and $\rho$ is the mass density of the material. For the numerical analysis of the experimental data we used several theoretical approaches, including the Parker, Cowan, and Clark - Taylor analysis. These analysis curves were plotted with the experimental temperature time rise in order to extrapolate diffusivity and correct the result for heat loss to side-walls of sample holder. The bismuth telluride films were prepared from the bulk $Bi_2Te_3$ crystals and stacked on a Corning glass substrate. The original $Bi_2Te_3$ crystal was used as a bulk reference. The laser-flash measurements revealed the RT thermal conductivity of the stacked films to be ~ 0.1-0.3 W/mK, which is significantly lower than that of the bulk reference sample ($K$=0.5 - 0.6 W/mK for cross-plane and $K$=1.5-2.0 W/mK for in-plane). Our results for the reference bulk $Bi_2Te_3$ crystal and thin films are in agreement with previously reported data for bulk and $Bi_2Te_3$-based nanoparticles [30, 48].

The second TPS technique (Hot Disk) provided $K$ values, which correspond to averaged in-plane component of the thermal conductivity. In the TPS measurements, a thin Ni heater-sensor covered with a thin electrically insulating layer was sandwiched between two stacked films under test. The samples were heated by short electrical current pulses for few seconds. The temperature rise in response to the dissipated heat was determined from the change in the resistance of the sensor. The time dependence of the temperature rise was used to extract the thermal conductivity from the equation [49] $\overline{\Delta T(\tau)} = P(\pi^{3/2} rK)^{-1} D(\tau)$, where $\tau$ is the parameter related to the thermal diffusivity $\alpha$ and the transient measurement time $t_m$ through the expression $\tau = \left(t_m \alpha / r^2\right)^{1/2}$, $r$ is the radius of the sensor, $P$ is the input power for heating the sample, and $D(\tau)$ is the modified Bessel function. The RT values obtained for thermal conductivity were ~1.1 W/mK, which also represent a significant drop compared to the bulk reference. The Seebeck coefficient was determined using MMR system (SB100) consisting of two pairs of thermocouples. One pair was formed with the junctions of copper and a reference material (constantan wire with the known Seebeck coefficient of ~ 36 μV/K). The other pair was formed with the junctions of copper and the layers under test. We modified the sample stage of the system in order to be able to use it with our thin films. The computer controlled sample stage was attached to the cold stage refrigerator and provides a pre-set stable temperature during the measurement. The sample chamber was kept at pressure below 10 mTorr while in operation. The measured value was about ~234 μV/m at RT.

***Acknowledgements***

The authors acknowledge the support from DARPA – SRC through the FCRP Center on Functional Engineered Nano Architectonics (FENA) and Interconnect Focus Center (IFC) as well as from US AFOSR through contract A9550-08-1-0100. Special thanks go to Mohammad Rahman and Zahid Hossain for their help with TEM and AFM measurements. The authors are indebted to other Nano-Device Laboratory (NDL) group members for assistance with the sample preparations.

**References**

[1] Ioffe, A.F., *Semiconductor Thermoelements* (Nauka, Moscow, **1956**) (in Russian); or Ioffe, A.F., *Semiconductor Thermoelectric and Thermoelectric Cooling* (Infosearch, London, **1957**.)

[2] Goldsmid, H.J.; Douglas, R.W. Thermoelectric *Br. J. Appl. Phys.* **1954**, *5*, 458.

[3] Wright, D.A. *Nature* **1958**, *181*, 834.

[4] Venkatasubramanian, R.; Siivola, E.; Colpitts, T.; O'Quinn, B.; *Nature* **2001**, *413*, 597-602.

[5] Xie, W. et al., *J. Applied Phys*. **2009**, *105*, 113713.

[6] DiSalvo *F.J. Science* **1999**, *285*, 703.

[7] Dresselhaus, M.S., et al. *Physics of the Solid State* **1999**, 41, 679.

[8] Hicks, L. D.; Dresselhaus, M. S. *Phys. Rev. B.* **1993**, *47*, 12727.

[9] Broido, D.A.; Reinecke, T.L. *Appl. Phys. Lett.* **1995**, *67*, 100.

[10] Balandin, A.;Wang, K.L. *Physical Review B* **1998**, *58*, 1544; Balandin A. and Wang K.L. *J. Applied Physics* **1998**, *84*, 6149.

[11] Zou, J.; Balandin, A.; *J. Applied Physics*, **2001**, *89*, 2932.

[12] Chen, G., Borca-Tasciuc, T., Yang, B., Song, D., Liu, W. L., Zeng, T., and Achimov, D. A. *Thermal Science and Engineering* **1999**, *7*, 43-51.

[13] Borca-Tasciuc, T., Achimov, D., Liu, W. L., Chen, G., Ren, H.-W., Lin, C.-H., and Pei, S. S. *Microscale Thermophysical Engineering* **2001,** *5*, 225-231.

[14] Feutelais, Y.; Legendre, B.; Rodier, N.; Agafonov, V. *Mater. Res. Bull.* **1993**, *28*, 591-596.

[15] For a review see Qi, X.-L.; Zhang, S.-C. *Physics Today*, January **2010**, 33-38.

[16] Bernevig, B.A.; Hughes, T.L.; Zhang, S.-C. *Science*, **2006**, *314*, 1757; Konig, M., et al., *Science*, **2007**, *318*, 766.

[17] Teweldebrhan, D.; Goyal, V.; Rahman, M.; Balandin, A.A. *Appl. Phys. Lett*., **2010**, *96*, 053107.

[18] Novoselov, K. S.; Geim, A. K.; Morozov, S. V.; Jiang, D.; Zhang, Y.; Dubonos, S.V.; Grigorieva, I. V.; and Firsov A. A. *Science*, **2004**, *306*, 666.

[19] Novoselov, K. S.; Jiang, D.; Schedin, F.; Booth, T. J.; Khotkevich, V. V.; Morozov, S. V.; Geim, A. K. *PNAS*, **2005**, *102*, 10453.

[20] Novoselov, K.S., Geim, A.K., Morozov, S.V., Jiang, D., Katsnelson, M.I., Grigorieva, I.V., Dubonos. S.V., and Firsov A. A. *Nature* **2005**, *438*, 197-200**.**

[21] Zhang, Y., Tan, J. W., Stormer, H. L. & Kim, P. *Nature* **2005**, *438*, 201–204.

[22] Calizo, I.; Balandin, A.A.; Bao, W.; Miao, F.; and Lau, C. N. *Nano Lett.* **2007**, *7*, 2645.

[23] Calizo, I.; Bao, W.; Miao, F.; Lau, C. N.; and Balandin A.A. *Appl. Phys. Lett.* **2007**, 201904.

[24] Calizo, I.; Bejenari, I; Rahman, M.; Liu, G. and Balandin, A.A. *J. Appl. Phys.* **2009**, *106*, 043509.

[25] Jones P.; Hubera,T. E., Melngailisb, J.; Barryb,J.; Ervin, M. H.; Zhelevac, T. S.; Nikolaevad, A.; Leonid Konopkod, Graf M. *IEEE Inter. Conf. Of Thermoelectrics* **2006**, 693.

[26] Golia, S.; Arora, M..; Sharma R.K.; Rastogi, A.C. *Current Appl. Phys.*, **2003**, 3, 195-197.

[27] Richter, W.; Kohler, H.; Becker, C. R. *Phys. Stat. Sol. B,* **1977**, *84*, 619.

[28] Russo V.; Bailini A.; Zamboni M.; Passoni M.; Conti C.; Casari C. S.; Li Bassi A.; and Bottani C. E. *J. Raman Spectroscopy*, **2008**, 39, 205-210.

[29] Lu W.; Ding Y.; Chen Y.; Wang Z. L.; Fang, J. *J. Am. Chem. Soc.* **2005**, *127*, 10112-10116.

[30] Cahill et. al; *J. Appl. Phys*. **2003** 93, 2; Chiritescu, C.; Mortensen, C.; Cahill, D.G.; Johnson, D.; Zschack, P. *J. Appl. Phys*., **2009**, *106*, 073503.

[31] Balandin, A.A.; Ghosh, S.; Bao, W.; Calizo, I.; Teweldebrhan, D.; Miao, F. and Lau, C.N.; *Nano Letters* **2008**, *8*, 902.; Ghosh, S.; Calizo, I.; Teweldebrhan, D.; Pokatilov, E.P.; Nika, D.L., Balandin, A.A.; Bao, W.; Miao, F.; and Lau, C.N. *Appl. Phys. Lett.* **2008**, *92*, 151911.

[32] Nika, D.L.; Ghosh, S.; Pokatilov, E.P.; Balandin, A.A. *Appl. Phys. Lett*. **2009,** *94*, 203103; Nika, D.L.; Pokatilov, E.P.; Askerov, A.S.; and Balandin, A.A. *Phys. Rev. B* **2009**, *79*, 155413.

[33] Lepri, S.; Livi, R.; Politi, A. *Phys. Rep.* **2003**, *377*, 1-80.

[34] Basile, G.; Bernardin, C.; Olla, S. *Phys. Rev. Lett.* **2006**, *96*, 204303.

[35] Dhar, A. *Phys. Rev. Lett.* **2001**, *86*, 5882.

[36] Ji; X.H., et al., *Materials Lett.*, **2005**, *59*, 682.

[37] Yang, J.Y; et al., *J. Alloys and Compounds* **2000**, *312*, 326.

[38] Shao, Q., Liu, G., Teweldebrhan, D.; Balandin, A.A.; Rumyantsev, S.; Shur, M. and Yan, D. *IEEE Electron Device Lett.* **2009,** *30*, 288.

[39] Liu, G.; Stillman, W.; Rumyantsev, S.; Shao, Q.; Shur, M.; and Balandin, A.A. *Appl. Phys. Lett.* **2009**, *95*, 033103.

[40] Teweldebrhan, D.; Balandin, A.A.; Methods for Producing Atomically-Thin Bismuth Telluride Films and Stacked Superlattices for Thermoelectric and Topological Insulator Applications, invention disclosure (University of California – Riverside, December 2009).

[41] Shamsa, M.; Liu, W.L.; Balandin, A.A.; Casiraghi, C.; Milne, W.I.; Ferrari, A.C. *Appl. Phys. Lett.*, **2006**, *89*, 161921.

[42] Ghosh, S.; Teweldebrhan, D.; Morales, J.R.; Garay, J.E.; Balandin, A.A. *J. Appl. Phys*., **2009**, *106*, 113507; Ikkawi, R; Amos, N.; Lavrenov, A.; Krichevsky, A.; Teweldebrhan, D.; Ghosh, S.; Balandin, A.A.; Litvinov, D.; Khizroev, S. *J. Nanoelectron. Optoelectron*. **2008**, *3*, 44-54.

[43] Shamsa, M.; Ghosh, S.; Calizo, I.; Ralchenko, V.; Popovich, A; Balandin, A.A.; *J. Appl. Phys.* **2008**, *103*, 083538.

[44] Balandin, A.A.; Shamsa, M.; Liu, W.L.; Casiraghi C.; Ferrari, A.C. *Appl. Phys. Lett.* **2008**, *93*, 043115.

[45] Cahill, D.; Watson, S.; Pohl, R. *Phys. Rev. B* **1992**, 46, 6131 – 6140.

[46] I. Bejenari; V. Kantser; A.A. Balandin, Phys. Rev. B. (in print) see also at arXiv: 0908.0624v2; Bejenari, I.; Kantser, V. *Phys. Rev. B* **2008**, *78*, 115322.

[47] Casian, A.; Sur, I.; Scherrer, H.; Dashevsky, Z. *Phys. Rev. B* **2000**, *61*, 15965.

[48] Dirmyer, M.R.; Martin, J.; Nolas, G.S.; Sen, A.; Badding, J.V., *Small*, **2009**, 5(8), 933-937.

[49] S. E. Gustafsson, Rev. Sci. Instrum. 62, 797 (1991).